\documentclass[%
 reprint,
superscriptaddress,
 amsmath,amssymb,
 aps,
pre,
]{revtex4-1}
\usepackage{enumerate}
\usepackage{graphicx}
\usepackage{dcolumn}
\usepackage{bm}
\usepackage{graphicx}
\usepackage{bbold} 
\usepackage{graphicx}
\usepackage{dcolumn}
\usepackage[left=3cm,top=2.5cm,right=3cm,bottom=2.5cm]{geometry}
\usepackage{array}
\usepackage{epsfig}
\usepackage{wrapfig}
\usepackage{color}
\usepackage{soul}
\usepackage{braket}
\usepackage{verbatim}
\usepackage{mathtools}
\usepackage{mathrsfs}
\usepackage{amsfonts}
\usepackage[english]{babel}
\usepackage{textcomp}
\usepackage{float}
\usepackage{comment}
\excludecomment{toexclude}
\usepackage{changepage} 

\begin{document}


\title{Evaluation of the path integral for flow through random porous media}


\author{Marise J. E. Westbroek}
\affiliation{Department of Earth Science and Engineering, Imperial College London, London SW7 2BP, United Kingdom}
\affiliation{The Blackett Laboratory, Imperial College London, London SW7 2AZ, United Kingdom}

\author{Gil-Arnaud Coche}
\affiliation{Accuracy, 41 Rue de Villiers, 92200 Neuilly-sur-Seine, France}

\author{Peter R. King}
\affiliation{Department of Earth Science and Engineering, Imperial College London, London SW7 2BP, United Kingdom}

\author{Dimitri D. Vvedensky}
\affiliation{The Blackett Laboratory, Imperial College London, London SW7 2AZ, United Kingdom}





\begin{abstract}
We present a path integral formulation of Darcy's equation in one dimension with random permeability described by a correlated multi-variate lognormal distribution.  This path integral is evaluated with the Markov chain Monte Carlo method to obtain pressure distributions, which are shown to agree with the solutions of the corresponding stochastic differential equation for Dirichlet and Neumann boundary conditions. The extension of our approach to flow through random media in two and three dimensions is discussed.
\end{abstract}

\pacs{Valid PACS appear here}
\maketitle

Flow through porous media \cite{sahimi93,ingham98,vafai15} is fundamental to many areas of science and engineering, including filtration \cite{zamani09}, casting and solidification \cite{stefanescu02}, physiology \cite{huang97,cook12}, hydrology \cite{kimura15}, and petroleum exploration and recovery \cite{taylor98}.   This has stimulated a wide range of theoretical and computational activity based on the Navier--Stokes equation \cite{stanley99}, lattice Boltzmann methods \cite{chen98}, cellular automata \cite{rothman88}, percolation theory \cite{hunt14}, and field theory \cite{king87}.

The are two basic approaches to calculating the flow through a porous medium.  In pore-scale simulations, fluid motion within the pore structure of the medium is obtained either directly from the Navier--Stokes equation, or indirectly from lattice Boltzmann simulations or cellular automata.  Although the microstructure of the pores in natural media, such as rocks, can be measured with X-ray tomography \cite{blunt17}, pore-scale simulations are not suitable for large regions because of their substantial computational requirements.

An alternative representation of the physics that enables larger regions to be studied is based on coarse-graining the medium and the flow over volumes large on the pore scale, but small on the scale of the macroscopic system.  The medium can then be described by an effective stochastic permeability of the pore structure. For the slow flow of incompressible viscous flow, the average flow rate and the average pressure gradient are related by Darcy's law:
\begin{equation}
{\bf q}=-K{\bf \nabla}p\, .
\label{eq1}
\end{equation}
Here, ${\bf q}$ is the average velocity of the fluid, $K=k({\bf x})/\mu$, where $k$  is the permeability of the medium, $\mu$ is the viscosity of the fluid, and ${\bf \nabla}p$ is the pressure gradient driving the fluid flow.  Although proposed as an empirical relation by Darcy in the mid-1850s, this equation is supported by homogenization of the Stokes flow \cite{whitaker86}, and for flow through a dilute random array of fixed obstacles \cite{rubinstein87}.

For an incompressible fluid, ${\bf \nabla}\cdot{\bf q}=0$, whereupon, with Darcy's law (\ref{eq1}), we obtain 
\begin{equation}
{\bf \nabla}\cdot(K{\bf \nabla}p)=0\, .
\end{equation}
In one dimension, the integration of this equation yields the simple form
\begin{equation}
{dp\over dx}=-{q_0\over K}\, ,
\label{eq3}
\end{equation}
where $q_0$ is a constant.

In this article, we formulate the solution of Darcy's equation for flow through a random permeable medium as a path integral which we then evaluate numerically using methods described elsewhere \cite{westbroek18} to obtain pressure statistics.  Path integrals provide a versatile formulation of Darcy's law by being amenable to direct numerical evaluation, as well as various approximations  \cite{amit84,zinn-justin06}.  Previous derivations \cite{drummond82,koplik94,teodorovich97} of path integrals related to Darcy's law have used their analytic properties to determine general flow characteristics.  Here, we compare the pressure statistics obtained from the path integral to the numerical integration of Darcy's law using the finite-volume method (FVM) \cite{schafer06} to demonstrate the accuracy of our methodology.

The path integral integral for Darcy's law is derived with methods used in classical statistical dynamics \cite{phythian77,peliti78,jouvet79,jensen81}.  A one-dimensional system of length $X$ is divided into $N_x$ segments of length $\delta x$. The pressure is defined at $0$, $(i-\frac{1}{2})\delta x$, for $i=1,2,\ldots N_x$, and $X$, so a stochastic pressure path through the system is $(p_0,p_1,\ldots,p_{N_x},p_{N_x+1})$.  The discrete form of Darcy's law (\ref{eq3}) on this lattice is
\begin{equation}
{p_i-p_{i-1}\over\delta x}=-q_0 e^{-L_i}\,.
\label{eq4}
\end{equation}
Here, $K_i=e^{-L_i}$ is the permeability at the $i$th grid point. The joint probability density of the log-permeabilities $L_i$ is taken as normal, which is a common assumption \cite{law44,li05}.

The stochastic generating functional for correlation functions of the pressure at fixed log-permeability is
\begin{align}
Z_{L}(\{u_i\})&=\int\prod_i dp_i  \exp\biggl(\sum_iu_ip_i\biggr)\nonumber\\
&\quad\times\delta\biggl({p_i-p_{i-1}\over\delta x}+q_0 e^{-L_i}\biggr)\, ,
\label{eq5}
\end{align}
The Jacobian $J=(\delta x)^{-N}$ from the argument of the $\delta$-function \cite{phythian77}, has been omitted.  Although $J$ becomes infinite as $\delta x\to0$, this quantity is cancelled by the same divergence in expressions for averages. Taking the average of $Z_L$ over the probability density of the log-permeability yields the generating function for pressure correlations:
\begin{equation}
Z(\{u_i\})=\int \prod_i dL_i P(\{L_i\})Z_{L}(\{u_i\}) e^{-\sum_i L_i }
\label{eq6}
\end{equation}
The factor $q_0$ in the Jacobian $q_0 \exp(-\sum_i L_i)$ has been omitted.
The log-permeabilities are taken to follow a Gaussian distribution:
\begin{align}
&P(\{L_i\})={1\over (2\pi)^{N/2}|{\sf C}_L|^{1/2}}\nonumber\\
&\quad\times\exp\biggl[-\sum_{ij}L_i({\sf C}_L^{-1})_{ij}L_j\biggr]\, ,
\label{eq7}
\end{align}
where ${\sf C}_L$ is the correlation matrix and $|{\sf C}_L|$ its determinant.  Substituting (\ref{eq5}) and (\ref{eq7}) into (\ref{eq6}) and again omitting constant prefactors,
\begin{align}
&Z(\{u_i\})=\int\prod_i dp_i\int\prod_i{d L_i}\exp\biggl(\sum_iu_ip_i\biggr)\nonumber\\
&\quad\times\exp\biggl[-\sum_{ij}L_i({\sf C}_L^{-1})_{ij}L_j \biggr]\exp\biggl[-\sum_i L_i\biggr]\nonumber\\
&\quad\times
\delta\biggl({p_i-p_{i-1}\over\delta x}+q_0 e^{-L_i}\biggr)\, ,
\end{align}
and integrating over the $L_i$, yields a path integral for the probability density $Q$ of the pressures:
\begin{equation}
Q(\{p_i\})={e^{-S(\{p_i\})}\over Z}\, ,
\label{eq9}
\end{equation}
where
\begin{equation}
Z=\int\prod_i dp_i e^{-S(\{p_i\})}\, ,
\end{equation}
with the discrete ``action''
\begin{align}
&S(\{p_i\})=\sum_i\log\biggl({p_{i-1}-p_i\over q_0\delta x}\biggr)\nonumber\\
&+\sum_{ij}\log\biggl({p_{i-1}-p_i\over q_0\delta x}\biggr)({\sf C}_L^{-1})_{ij}\log\biggl({p_{j-1}-p_j\over q_0\delta x}\biggr)\, .
\label{eq10}
\end{align}
Averages over pressure are determined by integrals over $Q$.  For example, the average $\langle p_k\rangle$ of the pressure $p_k$ at the $k$th lattice point is
\begin{equation}
\langle p_k\rangle={1\over Z}\int\prod_i dp_i\, p_k\, e^{-S(\{p_i\})}\, ,
\label{eq11}
\end{equation}
which confirms the cancellation of the omitted factors.   Higher-order correlation functions and cumulants are calculated analogously.

We use a Markov Chain Monte Carlo (MCMC) method to generate $N\gg1$ pressure paths representative of $Q$ in (\ref{eq9})--(\ref{eq11}).  For Dirichlet boundary conditions, $p_0=P_i$ and $p_{N_x+1}=P_f$ are fixed ($P_i>P_f$), while for Neumann boundary conditions, $p_0=P_i$ and $q_0$ are fixed.  The details of our implementation are described elsewhere \cite{westbroek18}, so we present only the salient steps here.  

Beginning with an initial pressure path, e.g., an array of random numbers that respects the boundary conditions, an update to a randomly selected pressure $p_i$ is accepted or rejected by its effect on the action according to the Metropolis--Hastings algorithm \cite{metropolis53,hastings70}:~acceptance with probability $\min\{1, e^{-\delta S}\}$ implies that proposed modifications that lower the action are always accepted.  Because each path is created from a previous path, there is strong autocorrelation within the Markov chain.  Therefore, a set of paths is representative of $Q$ only if a sufficient number of intermediate paths is discarded. Our calculations were done at different lattice spacings, to check their convergence. The calculation of autocorrelations and the verification of convergence are explained in Ref.~\cite{westbroek18}.

Both the path integral and the FVM require a distribution of permeabilities as input. Permeabilities distributed according to (\ref{eq7}) can be obtained from uncorrelated random numbers generated by an Ornstein--Uhlenbeck (OU) process \cite{gillespie96,chandrasekhar43}, whose general form (with zero mean) is
\begin{equation}
dL=-{L\over\xi}\,dx+\sigma\,\eta(x)\, dx ,
\label{eq13}
\end{equation}
where $\xi$ is a correlation length, $\sigma$ is a diffusion constant and $\eta$ is a stationary Gaussian process with mean zero, $\langle\eta(x)\rangle=0$, and correlation $\langle\eta(x)\eta(y)\rangle=\delta(x-y)$.  The solution to (\ref{eq13}) is
\begin{equation}
L(x)=L_0 e^{-x/\xi}+\sigma\int_0^x e^{-(x-y)/\xi}\eta(y)\,dy\, ,
\end{equation}
from which we calculate
\begin{align}
\langle L(x)\rangle&=L_0\,e^{-x/\xi}\, ,\\
\langle L(x)^2\rangle&=\textstyle{1\over2}\sigma^2\xi\Bigl(1-e^{-2x/\xi}\Bigr)\equiv\sigma^2\, ,\\
\hspace{-6pt}\langle L(x)L(y)\rangle&=\textstyle{1\over2}\sigma^2\xi\Bigl[e^{-|x-y|/\xi}-e^{-(x+y)/\xi}\Bigr]\, ,
\end{align}
with the notation $L_0\equiv L(0)$. As $x\to\infty$ the OU process approaches a stationary distribution with
\begin{equation}
\langle L(x)\rangle=0\, ,\quad\langle L(x)L(y)\rangle=\sigma^2 e^{-|x-y|/\xi}\, .
\end{equation}
Transcribing these results to the problem at hand, we define $L_i=\log K_i$, whereupon we have $\langle L_i\rangle=0$ (which implies that the geometric mean of the permeability is 1) and
\begin{align}
{\rm cov}(L_i L_j)&=\langle L_i L_j\rangle-\langle L_i\rangle\langle L_j\rangle\nonumber\\
&=\sigma^2 e^{-|i-j|/\xi}=({\sf C}_L)_{ij}\, ,
\label{eq19}
\end{align}
for the covariance of the log permeabilities.  

\begin{figure}[t]
\centering
\includegraphics[width=0.9\columnwidth]{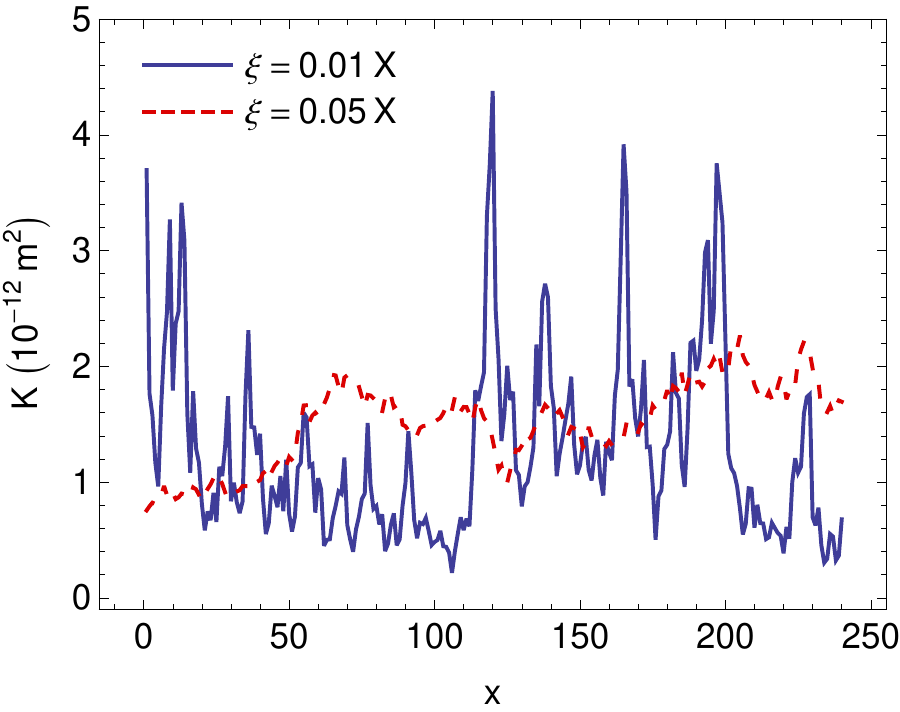}
\caption{(Color online) Two realizations of the variation of the permeability $K_i$ that are used in Darcy's law (\ref{eq4}) and in the path integral (\ref{eq7}).  For $\xi=0.01 X$ (solid blue line), the effect of correlations is  small, while, for $\xi=0.5 X$ (broken red line),  correlations throughout the system are evident.}
\label{fig1}
\end{figure}

Figure~\ref{fig1} shows the effect of the correlation length on the spatial variations of the permeability.  When $\xi=0.01 X$, the absence of correlations at longer distances is clearly evident in the large site-to-site fluctuations. However, when $\xi=0.5 X$ (half the system size), the permeability fluctuations are much smaller with a comparatively smoother profile.  Such differences in the permeability will be seen below to have a striking impact on the pressure profiles.

\begin{figure*}[t]
\centering
\includegraphics[width=12cm]{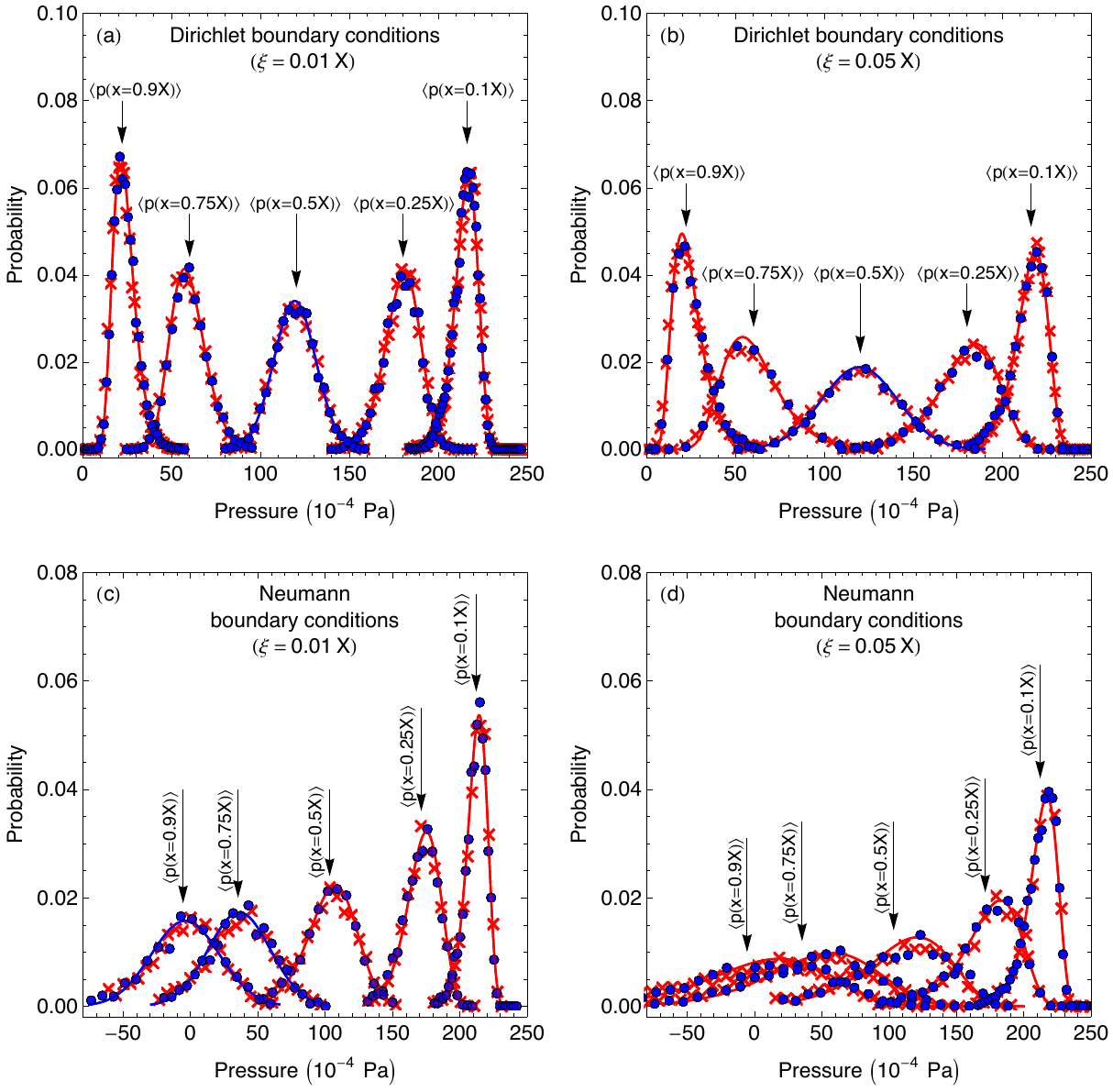}
\caption{(Color online) Pressure statistics obtained from the path integral (\ref{eq7}) (blue filled circles) and the FVM (red crosses) with (a,b) Dirichlet and (c,d) Neumann boundary conditions for (a,c) $\xi=0.01 X$ and (b,d) $\xi=0.05 X$.  Fewer data points are shown than used for statistical analysis.  Red curves are lognormal distributions (in the standard and reverse orientations) and blue curves are normal distributions, each with the same mean and variance as the pressures at the indicated positions.  Pressures are obtained from $10^4$ simulations on a system of 240 lattice sites with $\Delta x =0.5$~m by sampling the permeability distribution using (\ref{eq7}) and (\ref{eq19}). The sparse matrix solver UMFPACK \cite{UMFPACK} was used for the FVM simulations. To within statistical uncertainty (the error bars are smaller than the symbol sizes), the path integral and FVM simulations agree to a confidence level of 95\%.  The pressure gauge invariance of Darcy's equation means that only pressure differences are meaningful;~negative pressures are the result of a particular choice of a zero of pressure.}
\label{fig2}
\end{figure*}

Pressure distributions with Dirichlet and Neumann boundary conditions for the path integral (\ref{eq7}) and the FVM are shown in Fig.~\ref{fig2} for correlation lengths $\xi=0.01 X$ Fig.~\ref{fig2}(a,c) and  $\xi=0.05 X$ Fig.~\ref{fig2}(b,d) at five positions located symmetrically about $x=0.5X$. 
For Dirichlet boundary conditions, the pressure is fixed at each end of the system:~$p_0=P_i$ and $p_{N_x+1}=P_f$.   The average pressure is obtained by integrating Darcy's equation (\ref{eq3}) with these boundary values, 
\begin{equation}
\langle p(x)\rangle=\biggl(1-{x\over X}\biggr)P_i+{x\over X}P_f\, ,
\end{equation}
shows a linear decrease across the system with a slope of $(P_f-P_i)/X$.  Since $P_i>P_f$, the pressure distributions in Fig.~\ref{fig2}(a,b) are shown at positions that increase from right to left.  For Neumann boundary conditions, the pressure and flow are specified at $x=0$:~$p_0=P_i$ and $q_0=Q_0$.  The average pressure in the system is again obtained by integrating Darcy's equation (\ref{eq3}) with these boundary values, 
\begin{equation}
\langle p(x)\rangle=\biggl(1-{x\over X}\biggr)P_i+{x\over X}\langle p(X)\rangle\, ,
\end{equation}
which again shows a linear decrease across the system, so the pressure distributions in Fig.~\ref{fig2}(c,d) are also shown at positions that increase from right to left.  We have used a system size of $X=240~\mbox{m}$ with $P_i/X=2*10^4~\mbox{Pa/m}$ for all calculations.  For Dirichlet boundary conditions,  $P_f=0~\mbox{Pa}$ and $q=10^{-6}~\mbox{m/s}$, while, for Neumann boundary conditions, $Q_0=10^{-6}~\mbox{m/s}$.
These values reflect the application of Darcy's law to oil extraction from rocks. 

The most striking feature of Fig.~\ref{fig2} is the level of agreement between the two methods:~the Kolmogorov--Smirnov test \cite{KStest} indicates agreement at a confidence level of 95\%. This clearly demonstrates the accuracy of our evaluation of the path integral (\ref{eq7}).  In addition, there are several noteworthy characteristics of these pressure distributions.  For Dirichlet boundary conditions, the distributions are narrowest near the ends of the system, $x=0$ and $x=X$, where the pressure distributions are $\delta$-functions at the values specified by the boundary conditions.  The randomness of the permeability causes this distribution to broaden away from the endpoints.  Second, the pressure range is greater for the system with the smaller correlation length for the permeability.  This results from the pressure paths showing a smaller variation with the smaller correlation length.  With a larger correlation length, a small permeability persists over larger distances, resulting in suppressed pressures, while, for a smaller correlation length, a small permeability at one position can be followed by a large permeability at a nearby position. Finally, the pressure distributions are symmetric about $x=0.5X$.  This results from the permeability having been obtained from a stationary OU process \cite{pavliotis2014}. 

For Neumann boundary conditions, the pressure distribution at $x=0$ is a $\delta$-function at $P_0$. The width of the pressure distribution is, accordingly, narrowest near $x=0$, where the distribution is close to lognormal.  Unlike for Dirichlet boundary conditions, there is no fixed pressure at $x=X$ to cause subsequent narrowing of the distribution. Instead, the distribution continues to broaden and become more normal further into the system, with the lognormal distribution persisting further into the system for the larger correlation length.  A similar trend is seen for Dirichlet boundary conditions, with the distributions in the center of the system closest to normal.  These pressure distributions can be understood in terms of a random walk with a lognormal noise.  Regions separated by distances greater than the correlation length are effectively independent.  With an increasing accumulation of such regions, the central limit theorem dictates that the pressure statistics approach a normal distribution.

The comparisons in Fig.~\ref{fig2} demonstrate that the numerical evaluation of path integrals provides a viable alternative to the integration of Darcy's law with a stochastic permeability. However, there are differences in terms of runtime requirements. The FVM involves a matrix inversion for every pressure path. In one dimension, under Dirichlet boundary conditions, this $N_x\times N_x$ matrix is tridiagonal.  UMFPACK can solve a sparse matrix equation in $\mathcal{O}(N_x\log N_x)$ floating point operations (``flops'') \cite{UMFPACK}. The path integral requires $\mathcal{O}(N_x)$ flops to counter autocorrelations \cite{westbroek18}, bringing the total to 
$\mathcal{O}(N_x^{2})$ flops. However, there are techniques (e.g.~over-relaxation \cite{creutz87,brown87} and the multigrid method \cite{sokal86}) whose implementation is likely to decrease this run time considerably. 

In two and three dimensions, the matrices in the FVM are penta- and heptadiagonal, respectively. The run times generalize to $\mathcal{O}(N_x^d \log N_x^d)$ for the FVM and $\mathcal{O}(N_x^{2d})$ for the path integral. The higher-dimensional problem is further complicated by the lack of an analytic solution to Darcy's law. We will revisit this problem in a future publication. 

Finally, the analytic structure of the path integral suggests further study in relation to the randomness of the permeability.  For example,  coarse-graining the action may provide a route to determining how the distribution of the permeability changes as the viewing scale increases \cite{phythian77,peliti78,jouvet79,jensen81}.  Indeed, there has been a recent suggestion \cite{hansoge17} that the renormalization group, which has a natural formulation in terms of the path integral \cite{peliti78}, outperforms other upscaling techniques for Darcy's law in terms of accuracy and computational efficiency.

MJEW was supported by a Janet Watson scholarship from the Department of Earth Science and Engineering and a studentship in the Centre for Doctoral Training in Theory and Simulation of Materials (CDT in TSM), funded by the EPSRC (EP/L015579/1), both at Imperial College London. GAC was supported by a studentship in the CDT in TSM, funded by the EPSRC (EP/G036888/1), for the duration of his time at Imperial College London. We thank Grigoris Pavliotis for a helpful discussion.


\begin{thebibliography}{99}
\bibitem{NBCdiff} With Neumann boundary conditions, the matrix takes a slightly different form, but the essence of the problem is the same as with Dirichlet boundary conditions.
\end{thebibliography}
\end{document}